\documentclass[aps,reprint,twocolumn,superscriptaddress,citeautoscript,showpacs,amsmath,amssymb,floatfix,prl]{revtex4-1}
\usepackage{graphicx}
\usepackage{amsmath}
\usepackage{amssymb}
\usepackage{mathtools}
\usepackage{textcomp,gensymb} 
\usepackage{bm} 
\usepackage[usenames,dvipsnames]{xcolor}
\usepackage{float}
\usepackage{setspace}
\usepackage{dcolumn}
\usepackage{array}
\newcommand{\minus}{\scalebox{0.75}[1.0]{$-$}}

\usepackage{changepage}
\usepackage{tikz}
\usepackage{contour}
\usepackage[left]{lineno}
\usepackage{blindtext}
\usepackage{setspace}
\usepackage[colorlinks=true,citecolor=blue,linkcolor=blue,pdftex,hypertexnames=false,urlcolor=blue,linktocpage=true]{hyperref}
\definecolor{ashgray}{rgb}{0.7,0.75,0.71}
\definecolor{mspringgreen}{rgb}{0, 0.8, 0.1}
\definecolor{auburn}{rgb}{0.43, 0.21, 0.1}
\definecolor{ao(english)}{rgb}{0.0, 0.5, 0.0}
\definecolor{afw}{rgb}{0.95, 0.95, 0.96}
\definecolor{magnolia}{rgb}{0.97, 0.96, 1.0}
\definecolor{wsmk}{rgb}{0.96, 0.96, 0.96}

\newcommand{\ymb}{YbMnBi$_{2}$}

\newcommand{\cmb}{CaMnBi$_{2}$}

\newcommand{\emb}{EuMnBi$_{2}$}
\newcommand{\amb}{$A$MnBi$_{2}$}
\newcommand{\rmb}{$R$MnBi$_{2}$}

\newcolumntype{M}[1]{>{\centering\arraybackslash}m{#1}}

\newcommand{\RomanNumeralCaps}[1]
{\MakeUppercase{\romannumeral #1}}
\usepackage{dcolumn}

\usepackage{bbold}
\renewcommand{\vec}[1]{\bm{#1}}

\newcolumntype{d}[1]{D{.}{.}{#1}}

\begin{document}
\title{Signatures of coupling between spin waves and Dirac fermions in \ymb}
\author{A. Sapkota}
\email{sapkota@bnl.gov}
\affiliation {Condensed Matter Physics and Materials Science Division, Brookhaven National Laboratory, Upton, NY 11973, USA}
\author{L. Classen}
\email{lclassen@bnl.gov}
\affiliation {Condensed Matter Physics and Materials Science Division, Brookhaven National Laboratory, Upton, NY 11973, USA}
\affiliation {School of Physics and Astronomy, University of Minnesota,Minneapolis, MN 55415, USA}
\author{M. B. Stone}
\author{A. T. Savici}
\author{V. O. Garlea}
\affiliation{Neutron Scattering Division, Oak Ridge National Laboratory, Oak Ridge, TN 37831, USA}
\author{Aifeng Wang}
\altaffiliation{Present address: School of Physics, Chongqing University, Chongqing 400044, China}
\author{J. M. Tranquada}
\author{C. Petrovic}
\author{I. A. Zaliznyak}
\email{zaliznyak@bnl.gov}
\affiliation {Condensed Matter Physics and Materials Science Division, Brookhaven National Laboratory, Upton, NY 11973, USA}

\date{\today}

\begin{abstract}
We present inelastic neutron scattering (INS) measurements of magnetic excitations in \ymb, which reveal features consistent with a direct coupling of magnetic excitations to Dirac fermions. In contrast with the large broadening of magnetic spectra observed in antiferromagnetic metals such as the iron pnictides, here the spin waves exhibit a small but resolvable intrinsic width, consistent with our theoretical analysis.  
The subtle manifestation of spin-fermion coupling is a consequence of the Dirac nature of the conduction electrons, including the vanishing density of states near the Dirac points.  Accounting for the Dirac fermion dispersion specific to \ymb\ leads to particular signatures, such as the nearly wave-vector independent damping observed in the experiment.
\end{abstract}

\maketitle
Dirac and Weyl materials exhibit many exotic and novel quantum phenomena that are both of fundamental and potential technological interest \cite{Wehling_2014,Khodas_2009,Kefeng_2011,Aifeng_2016}. This class of materials encompasses a wide range of condensed matter systems including graphene, $d$-wave superconductors, and topological insulators and semimetals \cite{Wehling_2014}.
In these materials, the linear variation of energy as a function of wave vector about a Dirac node is a universal feature that leads to novel behaviors such as suppression of backscattering, high carrier mobility, impurity-induced resonant states, spin-polarized transport, and the unusual quantum Hall effect \cite{Wehling_2014,Khodas_2009,Kefeng_2011,Aifeng_2016,Zhang_2005,Hasan_2010}. Furthermore, interaction of these low-energy Dirac/Weyl fermions with other degrees of freedom leads to novel physics with technological potential \cite{Khodas_2009,Katmis_Nature2016,Masuda_2016}. Hence, understanding the coupling of Dirac fermions with other quantum excitations, such as spin waves, is a topic of great current interest.

From this perspective, 112 ternary bismuthides ($R,A$)MnBi$_2$ ($R =$ Rare-earth, $A =$ Alkaline-earth: Ca, Sr) represent a particularly interesting family where both magnetism and Dirac fermions coexist, providing a platform to study their interplay \cite{Kefeng_2011,Aifeng_2016,Park_2011,Zhang_2014,Guo_2014}. In these materials, the Dirac bands and the magnetic order are associated with distinct square-net layers: conducting Bi layers and magnetic MnBi layers separated by layers of $R,A$, 
as shown in Fig.~1(a) of Ref.~\citenum{Wang_2016}.
While indirect experimental evidence of a coupling between conduction electrons and magnetic Mn ions is provided by the impact of the magnetic order on electrical transport in \cmb\ \cite{Guo_2014,Wang_2012}, inelastic neutron scattering measurements on Sr/CaMnBi$_2$ found spin waves without any evidence of strong damping due to particle-hole excitations of the type seen in compounds such as CaFe$_2$As$_2$ \cite{Diallo_2009}.

\rmb\ systems were suggested as possible candidates where coupling of Dirac fermions with spins could be significant. \emb\ and \ymb\ are two such recently discovered materials \cite{Wang_2016}. \ymb\ is particularly interesting because of its possible link with  type-II 
Weyl physics with broken time-reversal symmetry \cite{Chinotti_2016,Borisenko_2015,Chaudhuri_2017}. In addition,
the ferromagnetic stacking of Mn moments along the $c$-axis, similar to \cmb, suggests that an interlayer exchange interaction can be mediated by Dirac bands \cite{Guo_2014,Rahn_2017}.
However, the questions remain: is there any theoretical signature of coupling/entanglement with Dirac fermions in the magnetic excitation spectrum of \ymb? If so, is it significant enough to be measured in a neutron-scattering experiment?

Here, we present the results of INS measurements performed on \ymb\ at $4$ different temperatures, spanning the N\'eel temperature, $T_{\rm N}$.  We show that the magnetic excitations are well-defined spin waves below $T_\mathrm{N}$, becoming dispersive paramagnons (similar to spin waves) just above $T_\mathrm{N}$, and the dispersion of both can be described with a local-moment Heisenberg model. In fitting the dynamical spin correlation function, we find that a good fit requires inclusion of a damping parameter $\gamma$, and that $\gamma$ has no significant variation with momentum transfer {\bf Q}.  While it is small, we note that no such damping is required in an insulating antiferromagnet such as CaMn$_2$Sb$_2$ \cite{McNally_PRB2015}. To understand the nature of damping from particle-hole excitations associated with the Dirac dispersion that has been experimentally reported for \ymb\ \cite{Borisenko_2015}, we have evaluated a spin-fermion coupling model in the framework of the random phase approximation (RPA). The results show that the damping should be small in our case due to the vanishing density of states near the Dirac points, which suppresses the effects of spin-fermion interaction, and effectively independent of {\bf Q}. Thus, our observed damping is consistent with a coupling to Dirac electrons and allows us to determine the coupling constant, which is directly proportional to the experimental damping parameter.


\begin{figure}[]
	\centering
	\includegraphics[trim = 27mm 115mm 65mm 98mm,clip,scale=0.72]{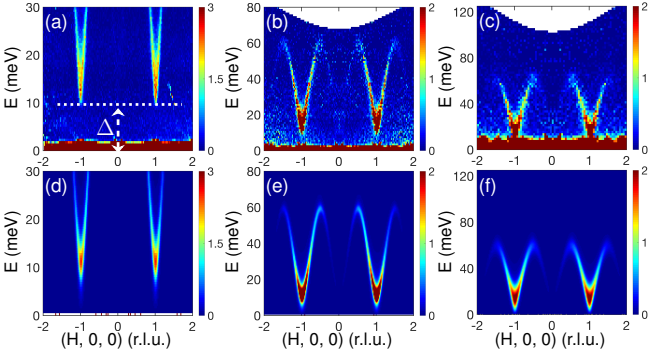}
	\caption{Spin waves in \ymb\ at $T = 4$~K. (a)--(c) INS spectra showing dispersion along $\left[H, 0, 0\right]$ direction measured with incident energies $E_\mathrm{i} = 35, 100$~and $200$~meV, respectively. Dashed line in (a) indicates the spin gap, $\Delta$ (cf Table~\ref{Tab2ddisp} and Fig.~S6). (d)--(f) INS spectra calculated using spin wave dispersion and Eqs.~\eqref{eq:S},\eqref{eq:DHO}, with the best fit parameters listed in Table~\ref{Tab2ddisp}. Intensity scales shown in the colorbars are in arbitrary units, which differ for different $E_\mathrm{i}$.}
	\label{Fig1}
	\vspace{-2mm}
\end{figure}
\begin{figure}[t]
	\centering
	\includegraphics[trim = 45mm 82mm 47.5mm 67mm,clip,scale=0.7]{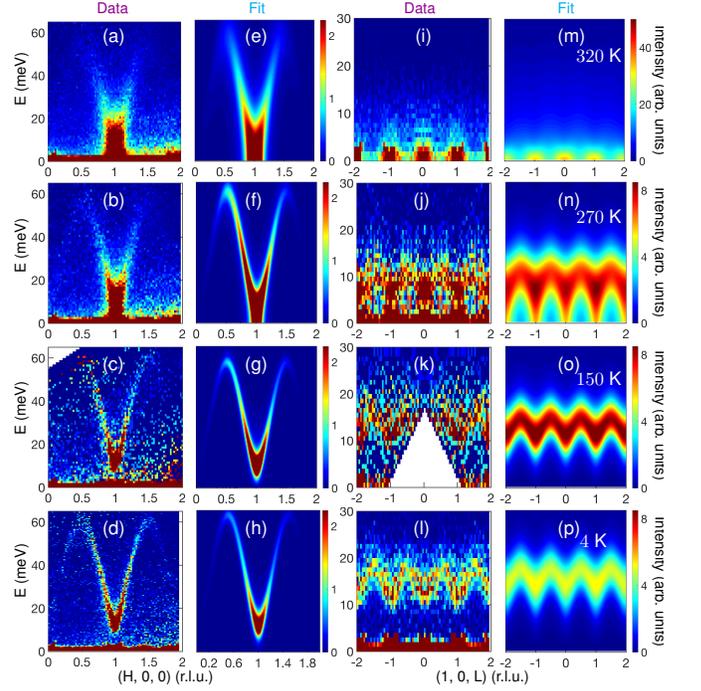}
	\caption{INS spectra measured on \ymb\ with $E_\mathrm{i} = 100$~meV at $T = 320$~K, $270$~K, $150$~K and $4$~K, illustrating the spin-wave dispersions along two symmetry directions, $\left[H,0,0\right]$ (a--d) and $\left[0,0,L\right]$ (i--l). (e--h) and (m--p), INS spectra calculated using Eqs.~\eqref{eq:S} and \eqref{eq:DHO} for the same wave vectors as the data, using the results of fits given in Table~\ref{Tab2ddisp}, and corrected for the instrument resolution (see \cite{Supplementary} for details).}
	\label{2ddisp}
	\vspace{-2mm}
\end{figure}
\begin{figure}[]
	\centering
	\includegraphics[trim = 43mm 112mm 44.5mm 96mm,clip,scale=0.7]{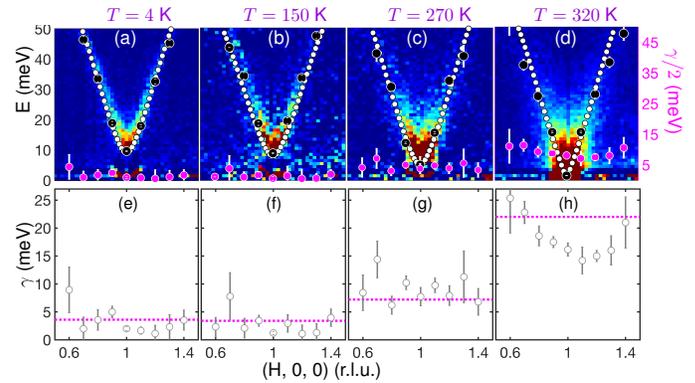}
	\caption{(a)--(d) The damping parameter $\gamma$ (magenta solid circles with white error bars) obtained from fitting 1D energy cuts using DHO response, and the low-energy part of spin wave dispersion at $T = 4$~K, $150$~K, $270$~K and $320$~K, respectively. Black circles show the corresponding spin wave energy, $E_{\textbf{q}}$, which was fitted for all temperatures. White circles illustrate the dispersion obtained using the parameters in Table~\ref{Tab2ddisp} for the corresponding temperature. The under-damped spin wave exists where $E_{\textbf{q}} > \gamma/2$. (e)--(h) Open circles are fitted values from 1D data and magenta dashed lines are for 2D data given in Table~\ref{Tab2ddisp}. Error bars show one standard deviation.}
	\label{corrw}
	\vspace{-2mm}
\end{figure}

Single crystals of \ymb\ were grown from Bi flux as described in Ref.~\citenum{Wang_2016}. \ymb\ orders antiferromagnetically below $T_\mathrm{N} \approx 290$~K, with an ordered moment of $4.3\mu_\mathrm{B}$ at 4~K \cite{Wang_2012,Wang_2016,Zaliznyak_2017}. INS measurements were performed at SEQUOIA (Figs.~\ref{Fig1}--\ref{corrw}) and HYSPEC \cite{Supplementary} spectrometers at the Spallation Neutron Source, Oak Ridge National Laboratory. Four single crystals with a total mass of $\approx 1.8$~g were co-aligned in the $(H, 0, L)$ horizontal scattering plane, with the effective mosaic spread of $\lesssim 0.8 \degree$ full-width at half-maximum (FWHM). The measurements were carried out with incident energies $E_i= 35$, 100 and 200~meV at $T=4$, 150, 270 and 320~K by rotating the sample about its vertical axis in $1\degree$ steps over a $270\degree$ range ($70\degree$ for $150$~K). Throughout the paper, we index $\textbf{Q} = (H, K, L)$ in reciprocal lattice units (rlu) of the P4$/nmm$ lattice, $a=b=4.48$~\AA, $c=10.8$~\AA\ \cite{Wang_2016}. {The collected event data were histogrammed into rectangular bins using the MANTID package \cite{MANTID}. For both the two-dimensional (2D) intensity maps (Figs.~\ref{Fig1}-\ref{corrw}) and one-dimensional (1D) cuts (see \cite{Supplementary}), the bin size of $\pm 0.0125$~rlu along $(H, 0, 0)$ and $\pm0.05$~rlu along $(0, K, 0)$ was used, except for $320$~K data, for which it was $\pm0.1$~rlu along $(0, K, 0)$; the bin size along $(0, 0, L)$ was $\pm0.1$~rlu.}

Figure~\ref{Fig1} (a)--(c) present INS spectra for \ymb\ in the antiferromagnetic (AFM) phase at $T = 4$~K, which reveal the spin wave dispersion along $\left[H, 0, 0\right]$ symmetry direction. Magnetic excitations are well-defined and sharp in both \textbf{Q} and $E$, indicating the presence of conventional spin-waves consistent with the local-moment description. The spin waves originate from the antiferromagnetic wavevector $\textbf{Q}_{\rm AFM} = (\pm1, 0, 0)$, as expected for a N\'eel-type magnetic order in \ymb\ \cite{Wang_2016}; the spin-gap $\Delta \sim 10$~meV marked in Fig.~\ref{Fig1}(a) is due to uniaxial anisotropy. The spin-wave dispersion bandwidth along $(H, 0, 0)$, $W = E_\mathrm{\textbf{Q} =(1.5,0,0)} - E_\mathrm{\textbf{Q} = (1,0,0)} \approx 50$~meV, is similar to the values measured in \amb\ \cite{Rahn_2017}.

To model the dispersion, we use a $J_{1}-J_{2}-J_{c}$ Heisenberg model, where $J_{1}$ and $J_{2}$  are the nearest- and next-nearest-neighbor in-plane exchange interactions, respectively, and $J_{c}$ is the inter-plane exchange. The dynamical spin correlation function, $S(\textbf{Q}, E)$, can be written as,
\begin{equation}\label{eq:S}
S(\textbf{q}+\textbf{Q}_{\rm AFM}, E) =
S_{\mathrm{eff}} \dfrac{(A_\mathrm{\textbf{q}} \minus B_\mathrm{\textbf{q}})}{E_\textbf{q}} \dfrac{f_\textbf{q}(E)}{1\minus e^{\minus E/k_{\mathrm{B}}T}} ,
\end{equation}
where $S_{\mathrm{eff}}$ is the effective fluctuating spin, $k_{\mathrm{B}}$ is the Boltzmann constant, and spin-wave theory gives $A_\mathrm{\textbf{q}} = 2S[2J_1 - 2J_2[\sin^2(\pi H) + \sin^2(\pi K)] - 2J_c\sin^2(\pi L) - D]$, $B_\mathrm{\textbf{q}} = 4SJ_1 \cos (\pi H) \cos(\pi K)$, and $E^2_{\textbf{q}} = A^2_\mathrm{\textbf{q}} - B^2_\mathrm{\textbf{q}}$. In the absence of damping, the spectral function, $f_\textbf{q}(E) = \delta(E-E_\textbf{q}) - \delta(E + E_\textbf{q})$, describes the conservation of energy. In the presence of damping, which we introduce to obtain a good fit to the full intensity distribution, delta-functions are replaced by Lorentzians \cite{Zaliznyak_1994} and the spectral function is given by the imaginary part of dynamical susceptibility of a damped harmonic oscillator (DHO),
\begin{equation}\label{eq:DHO}
  f_\textbf{q}(E) =  Z_\textbf{q} \dfrac{2 E_\textbf{q}}{\pi} \dfrac{\gamma E}{\left[E^2\minus E^2_{\textbf{q}}\right]^2 + (\gamma E)^2}.
\end{equation}
%
Here, $\gamma$ is the damping parameter (Lorentzian FWHM) and prefactor $Z_\textbf{q}$ ensures that the temperature-corrected DHO spectral function in Eq.~\eqref{eq:S} at all $\textbf{q}$ is normalized to 1 (for $(T, \gamma)\rightarrow 0$, $Z_\textbf{q} \rightarrow 1$) \cite{Supplementary}.

\begin{table*}[]
	\caption{Exchange coupling, uniaxial anisotropy, and damping parameters for \ymb\ obtained from fitting two-dimensional data shown in Fig.~\ref{2ddisp}. \label{Tab2ddisp}}
	\begin{ruledtabular}
		\begin{tabular}{cD{,}{\pm}{1.3} D{,}{\pm}{1.3} D{,}{\pm}{1.3} D{,}{\pm}{1.3}}
			\multicolumn{1}{c}{}&\multicolumn{1}{c}{$T = 4$~K\,\,\,\,}&\multicolumn{1}{c}{$T = 150$~K\,\,\,\,\,}& \multicolumn{1}{c}{$T = 270$~K\,\,\,\,\,} & \multicolumn{1}{c}{$T = 320$~K\,\,\,\,\,}\\
			\hline
			$SJ_1$ (meV) & 25.9 \, , \,0.2& 24.4 \, , \,0.3  & 27.1 \, , \,0.5& 25.6 \, , \,0.6\\
			$SJ_2/SJ_1$ & 0.39\, , \,0.01 & 0.40\, , \,0.01 & 0.43 \, , \,0.01& 0.41\, , \,0.01 \\
			$|SJ_c|/SJ_1$& 0.0050 \, , \,0.0001 & 0.0041  \, , \,0.0001 & 0.0022 \, , \,0.0001 & 0.0016 \, , \,0.0001\\
			$SD$ (meV) & -0.20 \, , \,0.01& -0.16 \, , \,0.01  & -0.06 \, , \,0.02& -0.003\, , \,0.001\\
			$\Delta$ (meV) & 9.1 \, , \,0.2& 8.0 \, , \,0.2 & 5.3\, , \, 0.8 & 1\, , \,1\\
			$\gamma$ (meV) & 3.6 \, , \,0.2 & 3.4 \, , \,0.2  & 7 \, , \,1 & 22 \, , \,4\\
		\end{tabular}
	\end{ruledtabular}
\end{table*}

We fit the data at each temperature using the cross-section given by Eqs.~\eqref{eq:S} and \eqref{eq:DHO} convolved with the instrumental resolution and accounting for the actual $({\bf Q},E)$ binning of the data. The apparent energy width is dominated by the wave vector resolution, which causes local averaging over the dispersion, so convolution with both the spectrometer resolution and binning functions is essential (additional details about the resolution correction and the fitting procedure are given in the Supplementary Materials \cite{Supplementary}). The resulting fits are shown side-by-side with the data in Figures \ref{Fig1} and \ref{2ddisp}, and the best fit values thus obtained are listed in Table~\ref{Tab2ddisp}. We find that the in-plane exchange parameters are nearly temperature-independent. The gap, $\Delta$, decreases with increasing temperature and approaches zero at $T \gtrsim T_\mathrm{N}$. 

Our central finding is a small but finite value of the damping parameter $\gamma$ at all temperatures (Table~\ref{Tab2ddisp}). It increases by roughly a factor of $6$ in the paramagnetic state, just above $T_\mathrm{N}$. While the global fits assume that $\gamma$ is independent of \textbf{Q}, we tested this assumption by fitting 1D energy cuts at different \textbf{Q} values, as shown in Fig.~\ref{corrw}; the corresponding plots of data with fits at 4~K and comparison calculations with $\gamma\sim 0$ are presented in Fig.~S4 \cite{Supplementary}. As one can see from Fig.~\ref{corrw}, $\gamma$ is non-negligible at all \textbf{Q} and with no systematic variation with \textbf{Q} (except possibly at $T>T_{\rm N}$). The values are generally within $2\sigma$ of the global fit values from Table~\ref{Tab2ddisp}, which are indicated in the bottom panels of Fig.~\ref{corrw} by magenta dashed lines.

In order to understand the possible origin of the spin wave damping, we first note that the contributions expected from magnon-magnon and magnon-defect scattering can be estimated by considering the case of the insulating antiferromagnetic layered compounds Rb$_2$MnF$_4$ \cite{Bayrakci_2013} and CaMn$_2$Sb$_2$ \cite{McNally_PRB2015}. In the first case, a high-resolution measurement of the magnon linewidths was performed, and the low-temperature ratio $\gamma/W$, was at least an order of magnitude smaller that our result; in the latter case, any damping was smaller than resolution. We also note that coupling to 2-magnon excitations cannot occur at energies below $2\Delta$, which is 18~meV in our case, and so such an effect is inconsistent with our observation of finite damping even at the lowest energies.  Thus, we conclude that simple magnon-magnon interactions do not provide a plausible explanation of our results.

The decay of magnons into electron-hole excitations can have a very large effect in itinerant magnets \cite{Diallo_2010,Aashish_cca_2017,Sapkota_2018,Jayasekara_2013}. The reason that similar effects are relatively small in our case is that unlike the parabolic dispersions of electrons in convention metallic systems, the linear dispersion of Dirac fermions results in a vanishing density of states near the Dirac nodes, which suppresses the available phase space for spin-fermion interactions. In addition, the Dirac fermions and spin waves in \ymb\ are spatially separated degrees of freedom that primarily propagate in different layers, which further inhibits the coupling. Hence, smaller broadening/damping with very different nature (both energy and \textbf{Q} dependence) than in metallic magnets should be expected.

\begin{figure}[]
	\centering
	\includegraphics[trim = 45mm 117mm 45mm 102mm,clip,width=0.47\textwidth]{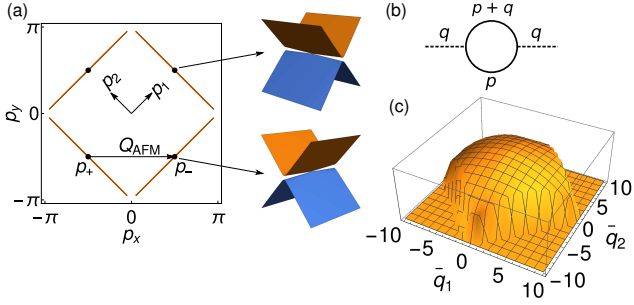}
	\caption{(a) The Fermi surface resulting from the anisotropic Dirac cones and a shift of the crossing away from the Fermi surface. The Dirac cones are rotated with respect to each other as sketched to the right. The centers of the ${\bf p}_{+}$ and ${\bf p}_{-}$ cones are connected by the antiferromagnetic wave vector ${\bf Q}_{\rm AFM}=(\pi,0)$. (b) Diagrammatic representation of the leading correction to the spin susceptibility through coupling to the Dirac electrons. Solid lines are Dirac propagators, dashed lines are spin waves. (c) The polarization Im$\Pi(E,\bar q_1, \bar q_2)$ obtained by numerical integration for an energy $E>\sqrt{2}v_\| q$ and $v_\|/v_\perp=0.005$ as function of $\bar q_i = v_\perp q_i$.}
	\label{fig:anDirac}
	\vspace{-2mm}
\end{figure}

Next, to establish that the spin-fermion coupling can indeed explain the observed non-neligible \textbf{Q}-independent damping, we calculate the correction to the spin susceptibility through coupling to the Dirac fermions within RPA and show that the observed damping is consistent with theoretical expectations for \ymb. We model the system of spins and conduction electrons in \ymb~via the action,
\begin{align}
S&= \int\! \!\!\frac{d^dp}{(2\pi)^d}\, \sum_{\eta=\pm}\left[  \psi^\dagger_{\eta}(p)  \left(i p_0 +\eta v_{1,\eta} \tau_x p_1 + v_{2,\eta} \tau_y p_2\right)\psi_\eta(p)\right] \notag \\ 
&+ \frac{g}{2} \int\!\! \!\frac{d^dp}{(2\pi)^d}  \int\!\!\! \frac{d^dq}{(2\pi)^d} \, \left[\vec{S}_{q} \psi^\dagger_{+}(p)  \vec{\sigma}\otimes \tau_x\psi_{-}(p-q) + \text{h.c} \right] \notag \\
&+ \int \!\!\!\frac{d^dq}{(2\pi)^d} \,\vec{S}_q \chi_0^{-1} \vec{S}_{-q}  \,,
\label{eq:action2}
\end{align}
which describes two Dirac cones at ${\bf p}_\eta$ [$\eta = \pm$, see Fig.~\ref{fig:anDirac}(a)] with anisotropic velocities, $v_{1,+} = v_{2,-} = v_\perp$ and $v_{1,-} = v_{2,+} = v_\|$, rotated with respect to each other ($v_\|/v_\perp\approx0.005$ in \ymb). We use rotated coordinates, $p_{1,2}=(p_x\pm p_y)/\sqrt{2}$. Pauli matrices ($\tau$) $\sigma$ act on (pseudo-)spin, respectively. The Mn spin waves are represented by the three-component boson field, $\vec{S}$. Their dynamical susceptibility is given by the bare expression, without damping, $\chi_0^{-1}(E)=E^2-E_\mathrm{\textbf{q}}^2$ [cf. Eq.~\eqref{eq:DHO}].

For simplicity, we perform calculations at $T = 0$ and in $d=2+1$ dimensions. We assume that the Dirac points have opposite chirality, as was found in the related compound, SrMnBi$_2$ \cite{Lee_2013}, and we consider a coupling, $\propto g$, which does not break chiral symmetry. In the coupling term in Eq.~\eqref{eq:action2}, we measure the wave vector transfer relative to the antiferromagnetic wave vector, ${\bf Q}_{\rm AFM}={\bf p}_{+}-{\bf p}_{-}$, which happens to connect the centers of the Dirac cones [see Fig.~\ref{fig:anDirac}(a)]. Because of the large anisotropy, $v_{\perp} \approx 9$~eV\AA\ and $v_{\|} \approx 0.043$~eV\AA\ in \ymb\ \cite{Chaudhuri_2017,Borisenko_2015}, the elliptical Fermi surface is extremely elongated and appears very similar to a true nodal line.

The leading correction to the bare susceptibility due to coupling to the Dirac electrons renormalizes the spin susceptibility via the polarization, $\chi^{-1}=\chi_0^{-1}-\Pi$, Fig.~\ref{fig:anDirac}(b). As the coupling is small and the semimetallic state in 2D Dirac materials is known to be stable due to a vanishing density of states at the Dirac points, we expect the second-order approximation for $\Pi$ to adequately capture the damping effects,
\begin{align}
\Pi(E,\vec{q})\!=\!-\frac{g^2}{2}\!\! \int\!\!\! \frac{d^dp}{(2\pi)^d} \text{Tr} [G_+(p_0,\vec{p})\tau_xG_-(p_0\!+\!E,\vec{p\!+\!q})\tau_x] \, ,
\end{align}
with Dirac propagators, $G_\eta(p_0,\vec{p})=(-ip_0+\eta v_{1,\eta}\tau_x p_1+v_{2,\eta}\tau_y p_2)/(p_0^2+v_{1,\eta}^2 p_1^2+v_{2,\eta}^2p_2^2)$. The damping can be determined by the imaginary part of the retarded polarization after analytical continuation, $ip_0\rightarrow E+i0^+$. For general $v_\perp,v_\|$, we obtain a lengthy expression, which can be found in Supplementary Materials \cite{Supplementary}.

It is instructive to consider two limits. For isotropic Dirac cones, $v_\perp=v_\|=v_F$, we find,
\begin{align}\label{eq:pi_vac}
\text{Im}\Pi^R(E,\vec{q})&=\frac{N_f}{8v_F^2}g^2 \text{sign}(E)\sqrt{E^2-v_F^2\vec{q}^2} \,\Theta(E^2-v_F^2\vec{q}^2) \,,
\end{align}
where $N_f$ is the number of Dirac cone \emph{pairs} and $\Theta$ is the step function. There are four Dirac points in each Brillouin zone of \ymb\ \cite{Chaudhuri_2017,Borisenko_2015}, so $N_f=2$. Although Eq.~\eqref{eq:pi_vac} has approximately the correct functional form for the DHO (Im$\Pi^R\approx c E$), the kinematic constraint, $E>v_F|\vec q|$, usually cannot be satisfied because for most wave vectors electronic energies are much larger than the spin-wave energy. For $E^2<v_F^2\vec{q}^2$, the polarization function is purely real \cite{Gonzalez_1996}.

The extreme anisotropy of the electronic dispersion in \ymb\ relaxes the kinematic constraint. 
For momentum transfers along ${\bf Q}_{\rm AFM}$, corresponding to the data shown in Figs.~\ref{Fig1}--\ref{corrw}, the leading order in small $(v_\|/v_\perp)$ reads,
\begin{align}
\text{Im}\Pi^R(E,q)\approx\frac{N_f}{2\pi v_\perp^2}g^2 E \, \Theta(E^2-2v_\|^2q^2) .
\end{align}
Accounting for further corrections leads to a weak momentum dependence \cite{Supplementary}. The full numerical result for Im$\Pi^R$ is presented in Fig.~\ref{fig:anDirac}. The main processes responsible for the enabled damping connect points along the elongated Fermi surfaces so that their energy cost is determined by $v_\|$. Due to its remarkable smallness, spin waves are able to excite such particle-hole pairs.

We conclude that the spin wave damping factor in Eq.~\eqref{eq:DHO} is given by $\gamma \approx N_fg^2/(2\pi v_\perp^2)$. Thus, using $\gamma \approx 3.6$~meV (Table~\ref{Tab2ddisp}), $N_f=2, v_F = 9$~eV\AA, we can estimate the coupling constant, $g \approx 1.0$~eV$^{3/2}$\AA. The obtained value of $g$ quantifies the spin-fermion interaction in \ymb\ and can be used, in future work, to analyze the effect of magnetism on the transport of Dirac electrons in the framework of Eq.~\eqref{eq:action2}.

In summary, we measured magnetic excitations in the Dirac material, \ymb, for temperatures in the range of $0.02 \leq {T}/{T_\mathrm{N}} \leq 1.10$. The results show dispersing spin waves for all temperatures and their detailed analysis unfolds the nature of spin-fermion coupling between the magnetic Mn layer and Dirac fermions of the Bi layer. We find a small, but distinct damping of spin waves, which for $T < T_\mathrm{N}$ is weakly dependent on temperature and is nearly independent of wave vector. Despite its small magnitude, the observed damping indicates substantial spin-fermion coupling parameter, $g \approx 1.0$~eV$^{3/2}$\AA, which we quantify by comparing the experiment with the theoretical analysis of model action for Dirac fermions coupled to spin waves, Eq.~\eqref{eq:action2}. Therefore, by combining the experimental measurements and theory, we establish the existence of long sought significant spin-fermion coupling in 112 family of Dirac materials.

\begin{acknowledgments}
We thank Alexei Tsvelik and Michael Scherer for helpful comments and discussions.
The work at Brookhaven was supported by the Office of Basic Energy Sciences, U.S. Department of Energy (DOE) under Contract No. DE-SC0012704. This research used resources at the Spallation Neutron Source, a DOE Office of Science User Facility operated by the Oak Ridge National Laboratory. LC acknowledges funding from a Feodor-Lynen research fellowship of the Humboldt foundation.
\end{acknowledgments}


%

\setcounter{page}{1}
\onecolumngrid
\section{\large{Supplemental Material:}}

\renewcommand{\thefigure}{S\arabic{figure}}
\setcounter{figure}{0}
\renewcommand{\thesection}{\Roman{section}}

\section{\romannumeral 1. Damped Harmonic Oscillator structure factor for Heisenberg spin model \label{HMDSHO}}

We analyzed the magnetic spectra using the $J_{1}-J_{2}-J_{c}$ Heisenberg model written as,
\begin{equation}
\label{hamiltonian}
H = J{_1}\sum_{\mathrm{NN}}\textbf{S}_i \cdot \textbf{S}_j+ J{_2}\sum_{\mathrm{NNN}}\textbf{S}_i \cdot \textbf{S}_j + J{_c}\sum_{\mathrm{NN(\perp)}}\textbf{S}_i \cdot \textbf{S}_j + D\sum_i(S_i^z )^2
\end{equation}
Here, $J_1$ and $J_2$ are the in-plane nearest-neighbor (NN) and next-nearest-neighbor (NNN) antiferromagnetic (AFM) exchange interactions and $J_c$ is the out-of-plane NN ferromagnetic (FM) interaction. The last term in Eq.~\eqref{hamiltonian} is the uniaxial anisotropy for the Mn spins corresponding to an easy axis along the $c$ direction, which is quantified by the anisotropy parameter $D < 0$. Using the Holstein-Primakoff representation of spin operators, in harmonic approximation the following spin-wave dispersion relation is obtained \cite{Mehmet_2017,Rahn_2017_SM},
\begin{equation} \label{disp1}
E_{\textbf{q}} = \sqrt{A^2_{\textbf{q}}-B^2_{\textbf{q}}} \, ,
\end{equation}
where $\textbf{q} = \textbf{Q}\minus \textbf{Q}_{\rm AFM} = (H, K, L)$ and,
\begin{eqnarray} \label{AqBq}
A_{\textbf{q}} &=& 4SJ_1 -2SJ_2[2-\cos (2\pi H)-\cos(2\pi K)] - 2SJ_c [1-\cos (2\pi L)] - 2SD \nonumber \\
&=& 4S[J_1 - J_2[\sin^2(\pi H) + \sin^2(\pi K)] - J_c\sin^2(\pi L) - D/2]\\
B_{\textbf{q}} &=& 2SJ_1[\cos(\pi H+\pi K)+\cos(\pi H-\pi K)] = 4SJ \cos(\pi H) \cos(\pi K)
\end{eqnarray}

The magnetic neutron scattering cross-section for a system of spins can be written as \cite{Squires_book_1978,ZaliznyakLee_MNSChapter,Sapkota_cca_2017_SM,Rahn_2017_SM},
\begin{equation}\label{NSC}
\dfrac{\text{d}^2\sigma}{\text{d}\Omega \text{d}E}\ = N r_m^2 |\frac{g}{2}F(\textbf{Q})|^{2}\dfrac{k_\mathrm{f}}{k_\mathrm{i}} \sum_{\alpha, \beta}(\delta_{\alpha\beta} \minus\hat{Q}_\alpha \hat{Q}_\beta)S^{\alpha \beta}(\textbf{Q}, E) \, ,
\end{equation}
where $N$ is the number of spins, $r_m = -5.39\cdot10^{-13}$~cm is magnetic scattering length, $g$ is Lande spectroscopic $g-$factor, $F(\textbf{Q})$ is the magnetic form factor (for \ymb, we use the magnetic form factor of Mn$^{2+}$ ion with an adjustable covalent compression \cite{Zaliznyak_2015} $p_{cov}$), $k_\mathrm{i}$ and $k_\mathrm{f}$ are initial and final neutron wave vectors, respectively, and $S^{\alpha \beta}(\textbf{Q}, E)$ is the dynamical structure factor describing spin-spin correlations. The dynamical structure factor is related to the imaginary part of dynamical spin susceptibility via the fluctuation-dissipation theorem,
\begin{equation}\label{FTSchi}
S(\textbf{Q}, E)=\dfrac{1}{\pi}\dfrac{\chi^{\prime\prime} (\textbf{Q},E)}{1\minus e^{\minus E/k_{\textrm{B}}T}} \, .
\end{equation}

The spectral function of a damped harmonic oscillator (DHO) accounts for the finite lifetime (non-zero inverse lifetime, $\Gamma=\gamma/2$) of a spin wave. The imaginary part of dynamical susceptibility for the DHO is \cite{Zaliznyak_1994_SM,LamsalMontfrooij_PRB2016},
\begin{equation}\label{DSHOchi}
\chi_\mathrm{DHO}^{\prime\prime}(E)=\dfrac{\gamma E}{(E^2\minus E^2_\mathrm{o})^2 + (\gamma E)^2} \, .
\end{equation}
Here, $E_\mathrm{o} = \hbar\omega_\mathrm{o}$, where $\omega_\mathrm{o}$ is the frequency of the undamped oscillator. In the underdamped regime, $E_\mathrm{o} > \Gamma = \gamma/2$, $\chi_\mathrm{DHO}^{\prime\prime}(E)$ can be represented via pole expansion, as a difference of the two Lorentzians \cite{Zaliznyak_1994_SM,LamsalMontfrooij_PRB2016},
\begin{equation}\label{UDHO}
\chi_\mathrm{DHO}^{\prime\prime}(E) = \dfrac{\pi}{2 \tilde{E}_\mathrm{o}}\left( \dfrac{1}{\pi}\dfrac{\Gamma}{(E - \tilde{E}_\mathrm{o})^2 + \Gamma^2} \minus \dfrac{1}{\pi}\dfrac{\Gamma}{(E + \tilde{E}_\mathrm{o})^2 + \Gamma^2} \right) \, ,
\end{equation}
where $\tilde{E}_\mathrm{o} = \hbar \tilde{\omega}_\mathrm{o} = \sqrt{E_\mathrm{o}^2 - \Gamma^2}$; $\tilde{\omega}_\mathrm{o}$ is the DHO frequency in the presence of damping.

To meet the requirement of the sum rule for the spin dynamical structure factor \cite{ZaliznyakLee_MNSChapter}, normalization of the DHO spectral function of Eq.~\eqref{FTSchi} has been implemented by calculating a prefactor, $Z_\textbf{q}^{-1} = \int_{-\infty}^{\infty}S(\textbf{q}, E)\text{d}E$, at each $\textbf{q}$. Using this normalization of $S(\textbf{Q}, E)$ with respect to energy and Eqs.~\eqref{FTSchi} and \eqref{DSHOchi}, we obtain the following dynamical structure factor for spin waves,
\begin{equation}\label{SDSHO}
\begin{split}
S(\textbf{q}+\textbf{Q}_\mathrm{N\acute{e}el}, E) = &
S_{\mathrm{eff}} \dfrac{2 Z_\textbf{q} }{\pi} \dfrac{(A_\mathrm{\textbf{q}} \minus B_\mathrm{\textbf{q}})}{1\minus e^{\minus E/k_{\mathrm{B}}T}}
\times \dfrac{\gamma E}{\big[E^2\minus E^2_{\textbf{q}}\big]^2 + (\gamma E)^2} \\
= & \dfrac{S_{\mathrm{eff}} }{1\minus e^{\minus E/k_{\mathrm{B}}T}} \times Z_\textbf{q} \dfrac{(A_\mathrm{\textbf{q}} \minus B_\mathrm{\textbf{q}})}{\tilde{E}_\textbf{q}} \left( \dfrac{1}{\pi}\dfrac{\Gamma}{(E - \tilde{E}_\textbf{q})^2 + \Gamma^2} \minus \dfrac{1}{\pi}\dfrac{\Gamma}{(E + \tilde{E}_\textbf{q})^2 + \Gamma^2} \right) \, ,
\end{split}
\end{equation}
where \textit{k}$_{\mathrm{B}}$ is the Boltzmann constant, \textit{S$_{\mathrm{eff}}$} is the effective fluctuating spin, $\gamma$ is the damping parameter corresponding to Lorentzian full width at half maximum (FWHM), $\Gamma = \gamma/2$ is the half width at half maximum (HWHM) and $\tilde{E}_\textbf{q} = \sqrt{E_\textbf{q}^2 - \Gamma^2}$ is the energy of a damped spin wave \cite{Zaliznyak_1994_SM}. In the limit of $\Gamma \rightarrow 0$, $Z_\textbf{q} \rightarrow 1$ and $\tilde{E}_\textbf{q} \rightarrow {E}_\textbf{q}$, and for $S_{\mathrm{eff}} = S$ Eq.~\eqref{SDSHO} yields delta function contributions describing an undamped linear spin wave \cite{Squires_book_1978}.

\section{\romannumeral 2. Spin wave dispersion from one-dimensional energy cuts \label{ODD}}

\begin{figure}[H]
	\centering
	\includegraphics[trim = 53mm 140mm 52.5mm 40mm,clip,scale=1.0]{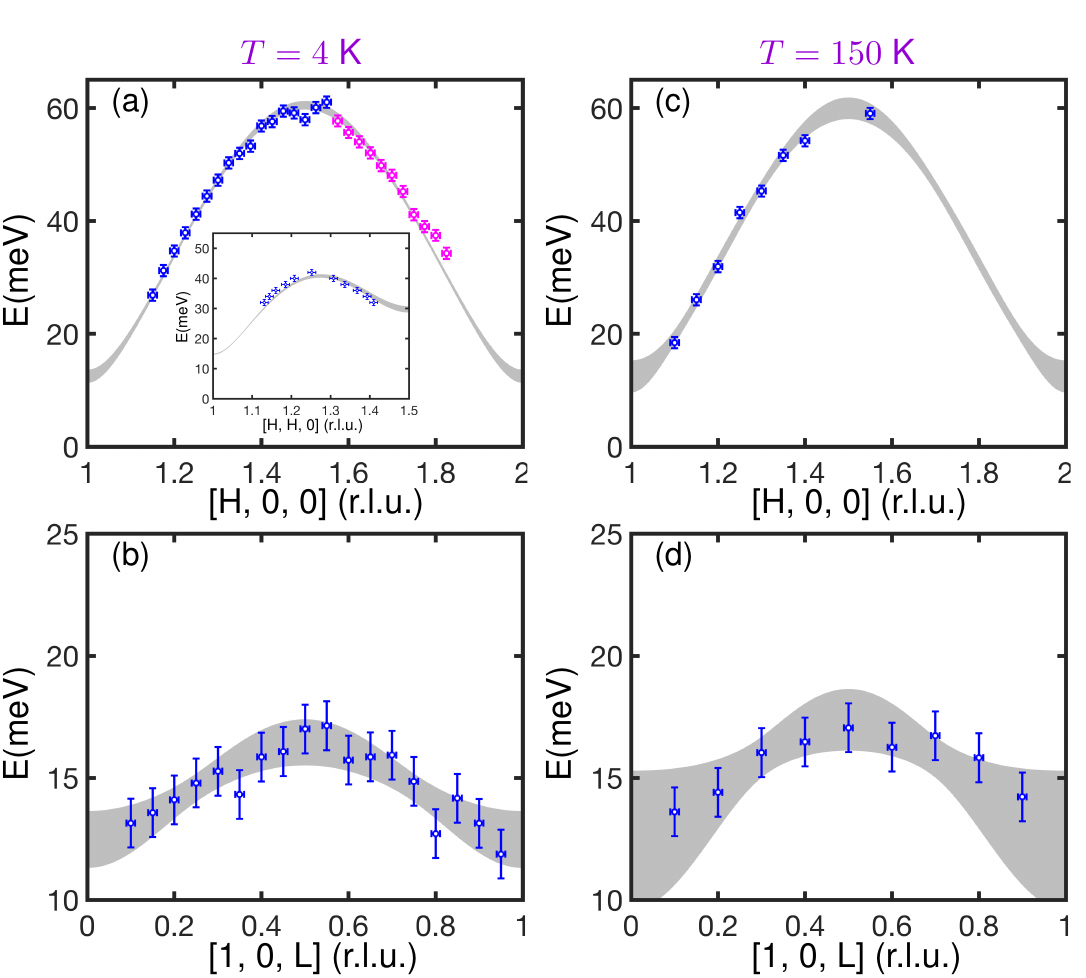}
	\caption{Spin-wave dispersions along high-symmetry directions, $\big[H, 0, 0\big]$ and $\big[1,0,L\big]$, for $4$~K (a, b) and $150$~K (c, d). The grey-shaded region represents the 95\% confidence interval of the fitting results given in Table~\ref{Tab1ddisp}. In (a), blue and magenta circles are obtained from two different $E$ vs $(H, 0, 0)$ dispersion data. Blue is from the data in which only $L = $~integers are averaged while the whole range of $ L = -6$ to $6$ were averaged for magenta. Only the blue data were used for the fit. The inset shows dispersion along the $\big[H, H, 0\big]$ direction. Error bars represent one standard deviation.}
	\label{1ddisp}
\end{figure}

One-dimensional dispersion data along $\big[H, 0, 0\big]$ and $\big[1, 0, L\big]$ directions at temperatures $4$~K and $150$~K shown in Fig.~\ref{1ddisp} were obtained using the peak position of one-dimensional (constant-\textbf{Q}) energy cuts of the two-dimensional (2D) magnetic scattering spectra shown in Fig.~\textcolor{blue}{$2$} of the main text. These 1D energy cuts were fitted using the Gaussian lineshape and accounting for Gaussian instrumental energy resolution. We note that this procedure does not account for the additional width, as well as the shift of the peak position measured near the bottom (top) of dispersion to higher (lower) energy, resulting from the convolution of spin wave dispersion with the wave vector resolution.

The peak positions thus obtained were used to create the dispersion plots shown in Fig.~\ref{1ddisp}, which are consistent with 2D plots of Figs.~\textcolor{blue}{$2$} and \textcolor{blue}{$3$} of the main text. The results obtained by fitting these 1D data with spin wave dispersion relation given by Eqs.~\eqref{disp1}--\eqref{AqBq} are shown in Table~\ref{Tab1ddisp}. They are fairly close to the results obtained from 2D fits presented in Table~\RomanNumeralCaps{1} of the main text. The underestimate of $J_1$ and overestimate of $\Delta$ in Table~\ref{Tab1ddisp}  result from the wave vector resolution effect mentioned above, thus providing an estimate of the magnitude of this effect. The wave vector resolution is properly accounted for, in the analysis presented in the main text, as described below.

\begin{table}[]
	\caption{Exchange and uniaxial anisotropy parameters and spin gap for \ymb\ obtained from fitting of the one-dimensional dispersion plots at $T = 4$~K and $150$~K shown in Fig.~\ref{1ddisp}.\label{Tab1ddisp}}
	\begin{ruledtabular}
		\begin{tabular}{cD{,}{\pm}{1.3} D{,}{\pm}{1.3}}
			\multicolumn{1}{c}{}&\multicolumn{1}{c}{$T = 4$~K\,\,\,\,}&\multicolumn{1}{c}{$T = 150$~K\,\,\,\,\,}\\
			\hline
			$SJ_1$ (meV) & 22.7 \, , \,0.3& 24\, , \,1 \\
			$SJ_2$ (meV) & 7.8 \, , \,0.2& 9 \, , \,1\\
			$SJ_c$ (meV) & -0.16 \, , \,0.03& -0.19 \, , \,0.06 \\
			$SD$ (meV) & -0.43 \, , \,0.04& -0.41 \, , \,0.07\\
			$\Delta$ (meV) & 12.4 \, , \,0.9& 12 \, , \,1\\
		\end{tabular}
	\end{ruledtabular}
\end{table}

\section{\romannumeral 3. Fitting procedure \label{FP}}

Prior to fitting the 2D data shown in the main text, we first fit 1D dispersions for $T = 4$~K and $150$~K shown in Fig.~\ref{1ddisp}. The fit parameters thus obtained (Table~\ref{Tab1ddisp}) are used as initial values for the fits of the 2D intensities shown in Fig.~\textcolor{blue}{$2$} of the main text. For higher temperatures, $270$~K and $320$~K, the values obtained from the fit of 2D data at $T = 4$~K are used for initial parameters.

For $4$~K and $150$~K, the data sets corresponding to the dispersions along $\big[H,0,0\big]$ and $\big[1,0,L\big]$ directions were fitted simultaneously. For $270$~K data, the in-plane exchange and the anisotropy parameters were first obtained by fitting the data along $\big[H,0,0\big]$ direction and then these parameters were fixed to the fitted values and $J_\mathrm{c}$ was obtained from the simultaneous fit of $(1,0,L)$ and $(H,0,0)$ data. For all the different fit procedures that we tried for $270$~K data, the fitted values of $\Gamma = \gamma/2$ were in the range $2.5-4.8$~meV and the average value is reported as the best fit parameter in the main text. For $300$~K, the paramagnon is overdamped near $(1, 0, 0)$ for $E < \Gamma$ and $(1,0,L)$ dispersion is not defined. Hence, the simultaneous fitting produces insensible imaginary results. We therefore obtained the values of the in-plane exchange, anisotropy, and damping parameters by fitting the data along $\big[H,0,0\big]$ and then estimated $J_\mathrm{c}$ from the $L-$dependent response of an over-damped oscillator by fitting the data along $\big[0,0,L\big]$ with these parameters fixed. All the fits were carried out with the magnetic form factor of Mn$^{2+}$, but using an adjustable covalent compression $p_{cov}$, $F(Q) = F_{Mn^{2+}} (p_{cov}Q)$, similar to Ref.~\citenum{Zaliznyak_2015}. The covalent compression was fitted for $4$~K data and was kept fixed to the refined value, $p_{cov} = 1.49(4)$, for data at higher temperatures.

\subsection{A. Account for the energy resolution\label{DCER}}

The fits were carried out accounting for both energy and wave vector resolution. Assuming that energy and wave vector resolution of the TOF spectrometer are decoupled \cite{Ehlers_2011, Abernathy_2012}, the DHO cross-section given by Eqs.~\eqref{NSC}--\eqref{SDSHO} was first convolved with a Gaussian lineshape in energy representing the instrumental energy resolution. In the under-damped case this simply replaces the Lorentzians in Eq.~\eqref{UDHO} with the corresponding Voigt functions. In order to avoid the contamination of the inelastic signal by the incoherent elastic background and magnetic Bragg peaks, only the data at $E > 7$~meV for $T = 4$~K and $150$~K and at $E > 5$~meV for $T = 270$~K and $320$~K were used for fitting.

\begin{figure}[]
	\centering
	\includegraphics[trim = 20mm 130mm 60mm 115mm,clip,scale=0.9]{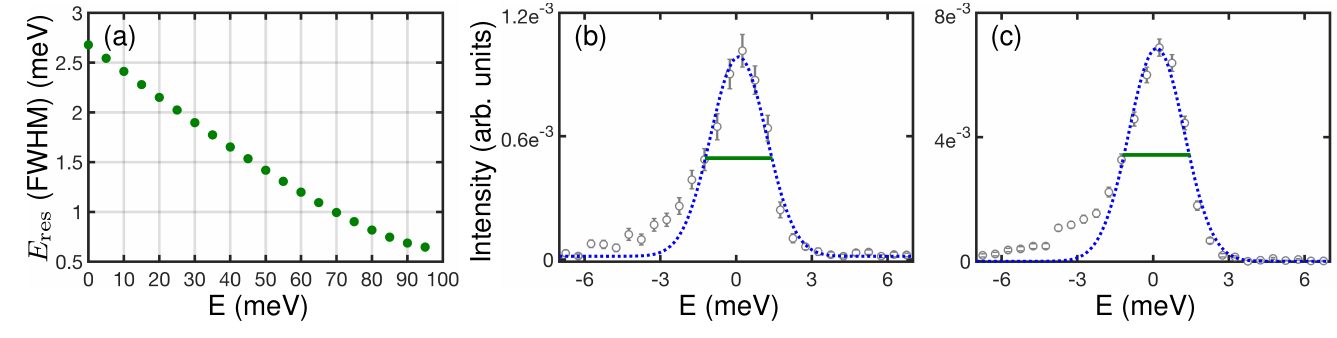}
	\caption{(a) Enegy resolution FWHM vs energy transfer for the incident energy $E_\mathrm{i} = 100$~meV and high-resolution SEQUOIA Fermi chopper spinning at $600$~Hz, corresponding to our measurement configuration. The $E_\mathrm{res}$ are calculated using the methods discussed in Refs.~\citenum{Ehlers_2011, Abernathy_2012}. (b) and (c) $E$-cuts at (0.45, 0, 0) and (1.45, 0 ,0), respectively, through $E = 0$~meV showing incoherent peaks. Cuts are taken from the data measured at $T = 4$~K. Blue dotted lines are fits to a Gaussian lineshape using only $ E > -2$~meV data and green solid lines represent the instrumental resolution width for $E = 0$~meV shown in panel (a).}
	\label{wcmp}
\end{figure}

The FWHM of the Gaussian energy resolution, $E_\mathrm{res}$, was calculated using the method discussed in Refs.~\citenum{Ehlers_2011, Abernathy_2012}. Fig.~\ref{wcmp} presents $E_\mathrm{res}$ for the high-resolution configuration with $E_\mathrm{i} = 100$~meV and SEQUOIA narrow-slotted Fermi-chopper \#2 spinning at 600 Hz used in our measurements. These values were used in convolution with the calculated cross-section for fitting the data. Fig.~\ref{wcmp} (b) and (c) illustrate that the calculated resolution at $E = 0$~meV very accurately reproduces the width of exemplary incoherent elastic scattering peaks at wave vectors (0.45, 0, 0) and (1.45, 0 ,0), respectively. This indicates that instrumental energy resolution is accurately accounted for in our analysis and the widths given in Table~\RomanNumeralCaps{1} of the main text represent the intrinsic broadening of the spin wave peaks.

\subsection{B. Account for the wave vector resolution\label{Qres}}

\begin{figure}[]
	\centering
	\includegraphics[trim = 60mm 113mm 65mm 97mm,clip,scale=0.8]{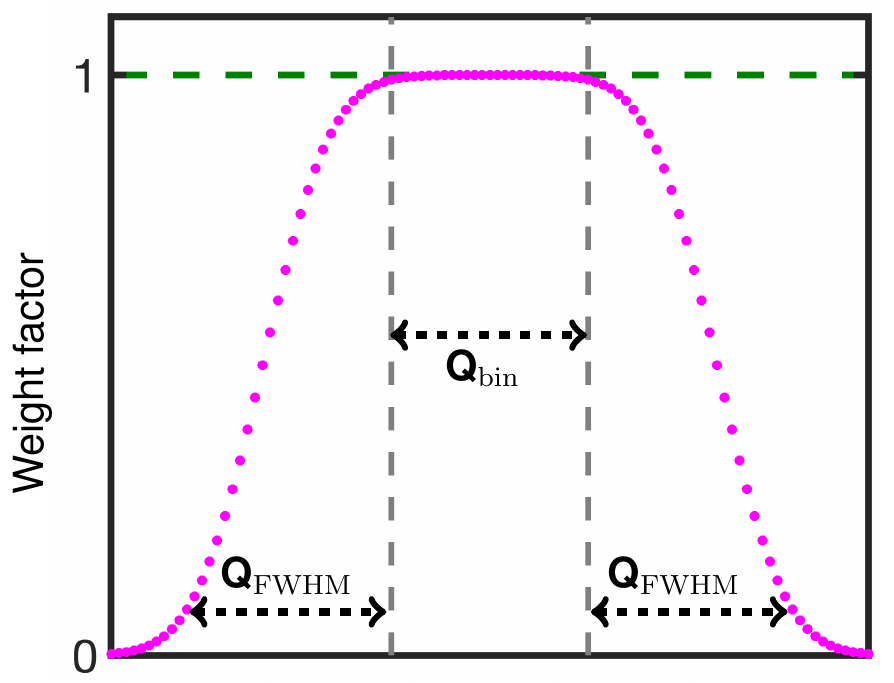}
	\caption{Illustration of the wave vector line shape used in our fits to account for the \textbf{Q} resolution. \textbf{Q}$_\mathrm{bin}$ is the range of the actual data bin used to create our plots and \textbf{Q}$_\mathrm{FWHM}$ is the calculated value of $Q$ resolution, using the methods discussed in Ref.~\citenum{Ehlers_2011}. In our fits of $2$D data, the spin wave intensity at each \textbf{Q} was obtained as a sum of intensities within the window of size \textbf{Q}$_\mathrm{bin} + 2$\textbf{Q}$_\mathrm{FWHM}$ weighted by the resolution weight function shown in the figure and normalized to 1, both centered at this \textbf{Q}.}
	\label{erff}
\end{figure}

\begin{figure}[]
	\centering
	\includegraphics[trim = 10mm 118mm 30mm 117mm,clip,scale=0.95]{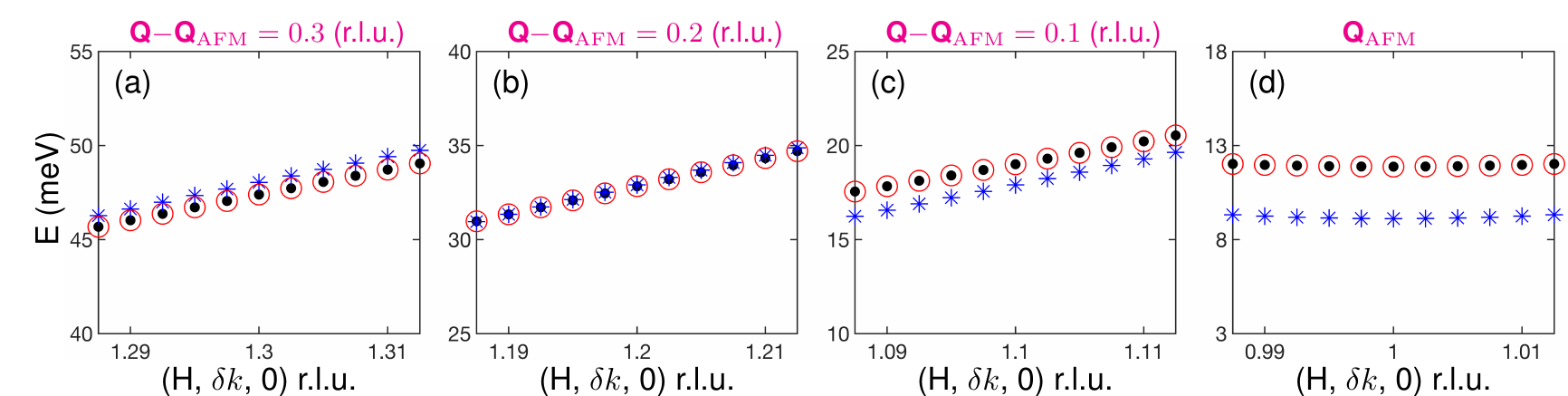}
	\caption{Illustration of the energy spans of the spin waves, at different \textbf{Q} along [H, 0, 0], corresponding to the binning $\delta h = \pm 0.0125$~r.l.u. and $\delta k = \pm 0.05$~r.l.u. (a)--(d) Markers are $E$ values obtained using the dispersion equations, Eqs.~\eqref{disp1}--\eqref{AqBq}, and $J$'s from Table~\RomanNumeralCaps{1} of the main text, at various \textbf{Q} $= (1.3/1.2/1.1/1.0, 0, 0) \pm (\delta h,\delta k,0)$, respectively. Markers with blue (asterisk), red (open circle), black (closed circle) colors are for $\delta k = 0$, and $\pm 0.05$~r.l.u., respectively. Energy spans of $\sim 5$~meV, corresponding to the \textbf{Q}-bin ($\delta h = \pm 0.0125$~r.l.u. and $\delta k = \pm 0.05$~r.l.u.) is also the energy range where the intensity above background exists in the binned INS spectra. Hence, the binning contributes $\lesssim 5$~meV to the peak width.}
	\label{esbw}
\end{figure}

As shown in Sec.~\textcolor{blue}{A} above, the energy resolution used in our high-resolution measurements is smaller than the intrinsic physical width of the spin wave peak extracted from our fits, which we assign to spin wave damping via decays into Dirac electron-hole pairs. The overall resolution correction to the width of the spin wave peak is also sensitive to the effect of wave vector resolution. The FWHM of the instrumental \textbf{Q} resolution was similarly calculated using the equations given in Ref.~\citenum{Ehlers_2011} and accounting for the sample mosaic of $0.8^\circ$. The average value of \textbf{Q}$_\mathrm{FWHM}$ in the $5$~meV energy transfer window was used in our fits.

In order to account for the additional wave vector broadening due to the binning of the data at each \textbf{Q}, the Gaussian wave vector resolution was convoluted with the window function corresponding to the wave vector range used for binning. This results in a \textbf{Q} resolution function in the form of two complementary error functions parameterized by the \textbf{Q}-resolution FWHM, \textbf{Q}$_\mathrm{FWHM}$, and the bin size, \textbf{Q}$_\mathrm{bin}$, as shown in Fig.~\ref{erff}. In order to minimize the effect of wave vector resolution and still have sufficient intensity for reliable fitting, we used the optimized bin size of $\pm 0.0125$ and $\pm0.05$ along $(H, 0, 0)$ and $(0, K, 0)$, respectively, for all the $E_\mathrm{i} = 100$~meV and $E_\mathrm{i} = 35$~meV data, except for $320$~K data, for which $\pm0.1$ along $(0, K, 0)$ direction was used; the bin size in $(0, 0, L)$ was kept at $\pm0.1$.
{A crude estimate using the spin-wave dispersion slope of $\sim 120$ meV/rlu along the \big[H, 0, 0 \big] direction shows that bin size $\delta h = \pm 0.0125$~rlu contributes $\sim 3$~meV to the peak energy width. This is illustrated in Figure~\ref{esbw}, which shows that in the absence of any intrinsic width the binning ($\delta h = \pm 0.0125$ and $\delta k = \pm 0.05$) would introduce peak width of $\sim 5$~meV for all energies and wave vectors along \big[H, 0, 0\big]. This is comparable to the values of $\gamma$ obtained from our analysis and therefore it is important that analysis (Fig.~\ref{erff}) accounts for this artifact, which exists in the data where the width introduced by binning is significant}.

\subsection{C. The manifestation of the intrinsic physical width in the line shape of spin wave peak\label{DNNW}}

\begin{figure}[b]
	\centering
	\includegraphics[trim = 54mm 89mm 56mm 73mm,clip,scale=0.85]{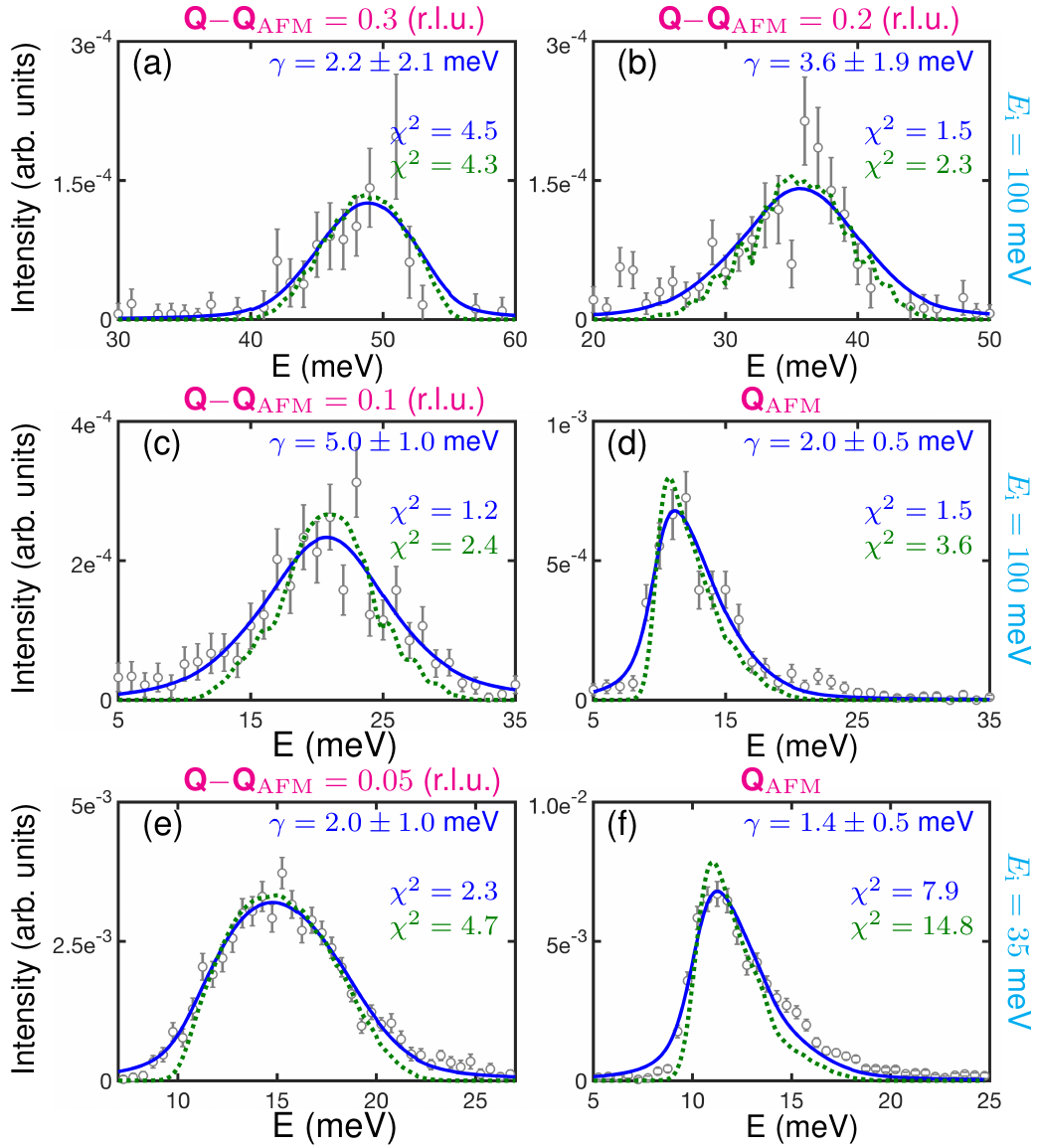}
	\caption{Selected one-dimensional $E$-cuts of $T = 4$~K data with resolution-corrected fits to Eq.~\eqref{SDSHO}. (a)--(d) cuts at four ~\textbf{Q} points along the $\big[H, 0, 0\big]$ direction for the data measured with $E_\mathrm{i} = 100$~meV. (e) and (f) cuts at two ~\textbf{Q} points near the bottom of the dispersion for the data measured with $E_\mathrm{i} = 35$~meV. Blue solid lines are the DHO fits, Eq.~\eqref{SDSHO}, where $\gamma$ was varied and green dotted lines are fits with fixed $\gamma = 0.02$~meV. Fitted values of $\gamma$ obtained for the blue lines are given in each figure. Data are averaged over the range of $\pm 0.0125$ and $\pm0.05$ along $\big[H, 0, 0\big]$ and $\big[0, K, 0\big]$, respectively. Error bars in all figures represent one standard deviation.}
	\label{DPNW}
\end{figure}

Figures~\ref{DPNW} (a)--(f) present best fits of the one-dimensional energy cuts of the 4~K data, where the fits are done considering the instrumental resolution and bin width, as discussed above in Section~\textcolor{blue}{A} and~\textcolor{blue}{B}. In addition, the fits were done with and without accounting the intrinsic peak width, $\gamma$. The results demonstrate the consistency of non-negligible broadening with the measured width of the spin waves. Green dotted lines are fits with $\gamma=2\Gamma = 0.02$~meV and blue lines are fits with free $\gamma$. The $\chi^{2}$ corresponding to the green lines with negligible damping is systematically larger than that of blue lines where $\gamma$ was fit. Green lines are also clearly less accurate in describing the broad tail of the spin wave peak, which is especially apparent in panels (c) -- (f). 
The difference is clear, indicating that intrinsic width, $\gamma \gtrsim 2$~meV, is indeed present even at the lowest temperature.

\section{\romannumeral 4. General expression of polarization bubble}

\begin{figure}[b]
	\centering
	\includegraphics[width=0.3\textwidth]{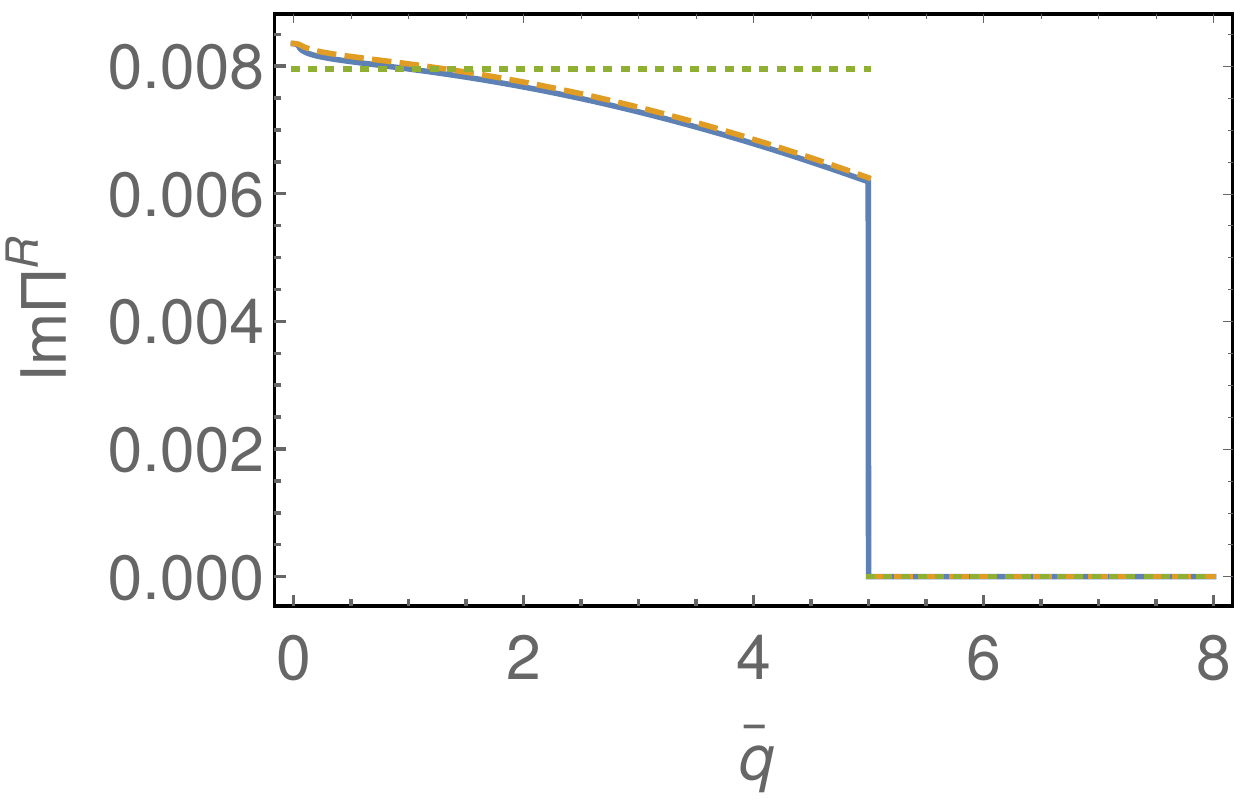}\quad
	\includegraphics[width=0.3\textwidth]{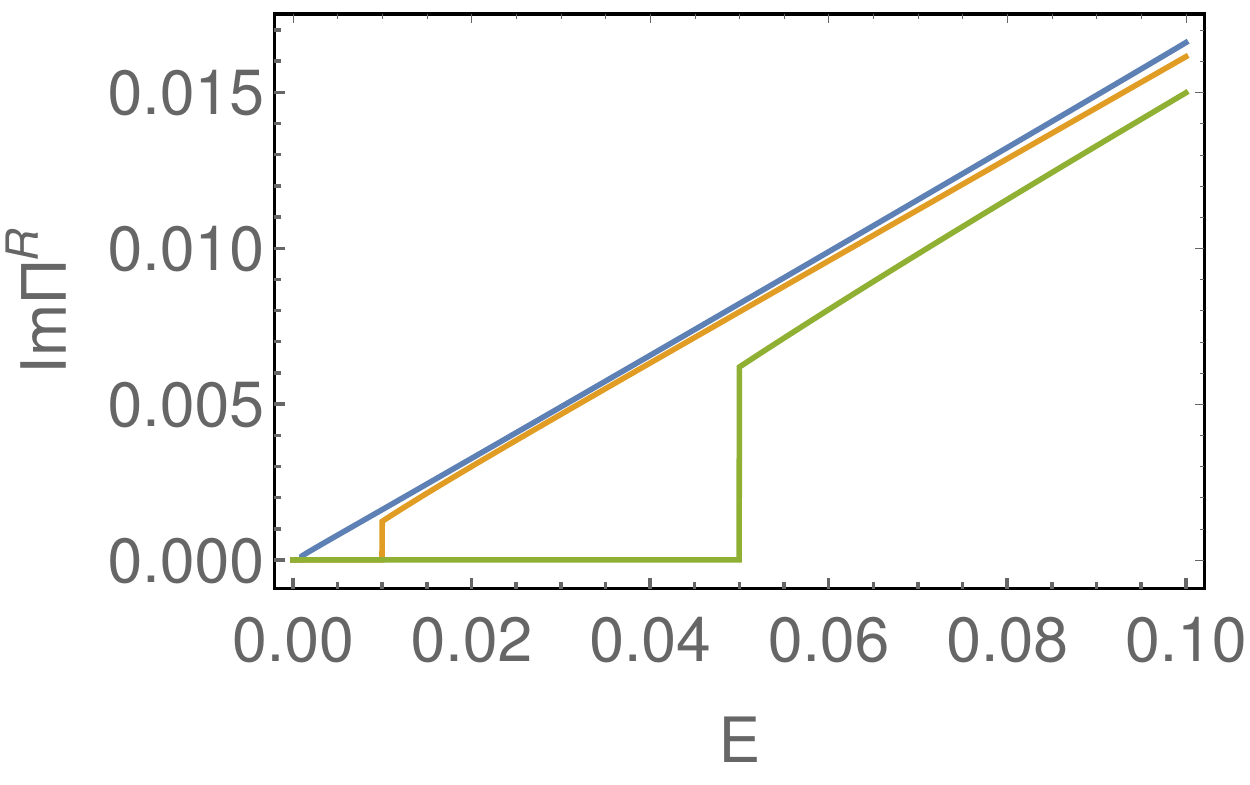}\quad	
	\includegraphics[width=0.3\textwidth]{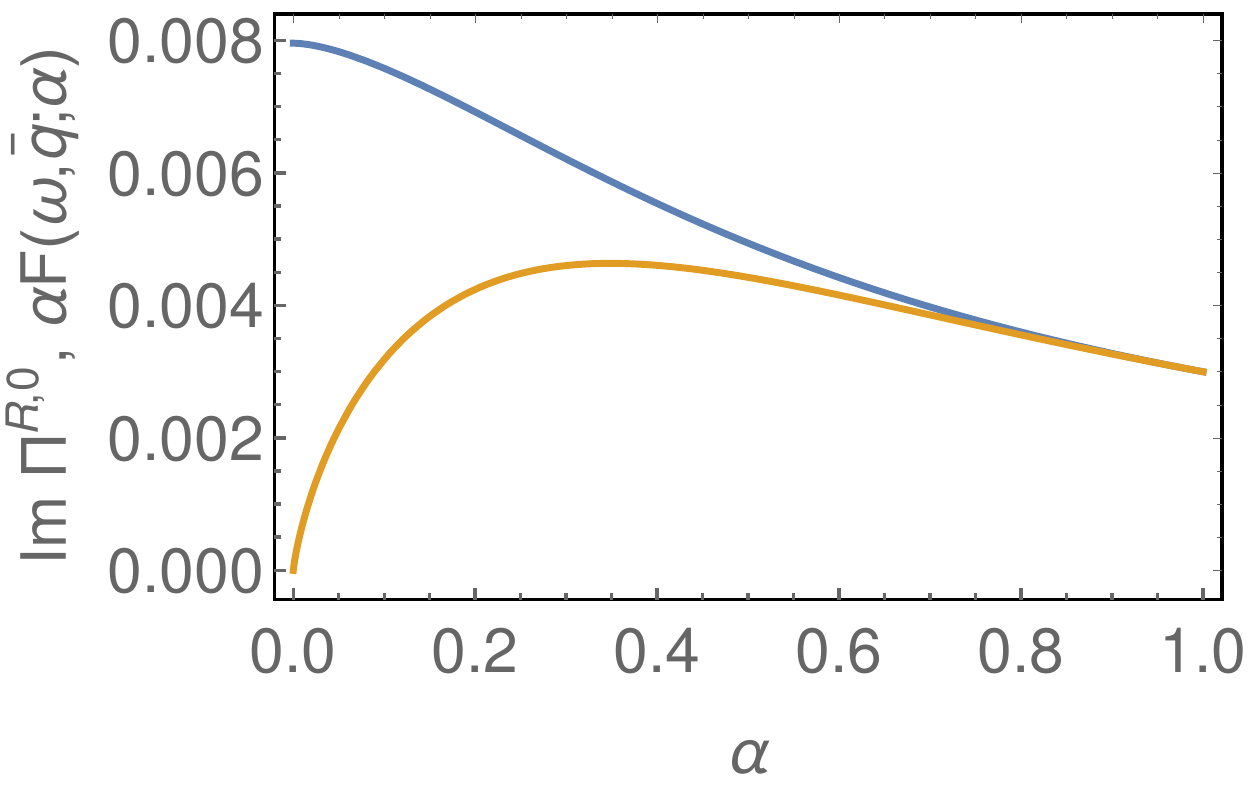}
	\caption{(left) Polarization as function of momentum transfer for fixed energy $E=50$meV and anisotropy $\alpha=0.005$. We show different approximations: full numerical expression, Eq.~\eqref{eq:anPi}, (solid), expansion in $\alpha$ with corrections, Eq.~\eqref{eq:approx1}, (dashed), and only the leading term, Eq.~\eqref{eq:approx0}, which is momentum-independent (dotted). The kinematic constraint, i.e. the location where the damping vanishes, is well reproduced in all cases. (middle) Polarization as function of energy for different momentum transfers, $v_\perp q = 0.1, 1,$ and $5$~eV. For energies that are larger than the threshold required to excite a particle-hole pair, $E>v_\| q$, the polarization for a fixed momentum is a linear function of energy, as described by Eq.~\eqref{eq:approx0}. The vertical displacement can be estimated as $\frac{N_f}{8\pi v_\perp^2}g^2 \alpha F(0,q)$ with $\alpha F(0,q)\approx 4\alpha \bar q$. (right) Numerically calculated $\alpha F(\omega,\bar q, \alpha)$ (orange) and leading contribution to the polarization, Im$\Pi^{R,0}=\text{Im}\Pi^R-\alpha F$ (blue), as function of $\alpha$ for $\omega=50$meV and $\bar q=10$meV. For the experimental value of $\alpha$ in YbMnBi$_2$, $\alpha\approx 0.005$, the correction is negligible. }
	\label{fig:anPi}
\end{figure}

With the coupling between Dirac electrons and spin as given in the main text, we obtain within RPA for the polarization
\begin{align}
\text{Im}\Pi^R(E,\vec{q})=-\frac{N_f}{4\pi v_\perp^2}g^2 \text{sign}(E) \int_0^1 &dx \,\Theta\left(E^2-\alpha^2\left( \frac{\bar q_1^2}{1-x+\alpha^2 x} +  \frac{\bar q_2^2}{x+\alpha^2(1- x)} \right)\right) \frac{1}{\sqrt{(1-x+\alpha^2 x)(x+\alpha^2(1- x))}} \notag \\
&\times \left[-\sqrt{|\Delta_2|} + \frac{\Delta_1}{\sqrt{|\Delta_2|}} -\alpha \frac{1+\alpha^2}{(1-x+\alpha^2 x)(x+\alpha^2(1- x))} \sqrt{|\Delta_2|} \right]
\label{eq:anPi}
\end{align}
with
\begin{align}
\Delta_2&=x(1-x)\left[ -E^2 + \alpha^2\left( \frac{\bar q_1^2}{1-x+\alpha^2 x} +  \frac{\bar q_2^2}{x+\alpha^2(1- x)} \right) \right] \\
\Delta_1&=x(1-x)\left[ -E^2 + \alpha^3\left( \frac{\bar q_1^2}{(1-x+\alpha^2 x)^2} +  \frac{\bar q_2^2}{(x+\alpha^2(1- x))^2} \right) \right]
\end{align}
and $\alpha=v_\|/v_\perp$, $\bar q_i= v_1 q_i$, i.e. $\alpha \bar q_i = v_\| q_i$. We also introduced a Feynman parameter, $x$. Expanding in small $\alpha$, we see that in this case the kinematic constraint is indeed relaxed, giving the expected damping. Let us consider momentum transfers along the antiferromagnetic wave vector, which corresponds to $q_1=q_2=q/\sqrt{2}$. The leading order then reads,
\begin{align}
\text{Im}\Pi^R(E,q)\approx\frac{N_f}{2\pi v_\perp^2}g^2 E \, \Theta(E^2-2v_\|^2q^2) ,
\label{eq:approx0}
\end{align}
as given in the main text. Accounting for small corrections, we find,
\begin{align}
\text{Im}\Pi^R(\omega,\bar q)\approx-\frac{N_f}{4\pi v_\perp^2}g^2 \text{sign}(\omega) \Theta(\omega^2-2\alpha^2\bar q^2) \left[-2\omega E\left(1-\frac{2\alpha\bar q^2}{\omega^2}\right) + \frac{2\alpha^2 \bar q^2}{\omega^2}K\left( 1-\frac{2\alpha\bar q^2}{\omega^2} \right) \right] + \frac{N_f}{8\pi v_\perp^2}g^2\alpha F(\omega,q;\alpha)\,,
\label{eq:approx1}
\end{align}
where $K(x)$ and $E(x)$ are the complete elliptic functions of first and second kind, and
\begin{equation}
F(\omega,\bar q;\alpha)=-\int_0^1 dx\Theta\left(\omega^2 x(1-x)-\alpha^2 \bar q^2/2\right) \frac{1+\alpha^2}{\sqrt{(1-x+\alpha^2 x)(x+\alpha^2(1- x))}^3} \sqrt{|\omega^2 x(1-x)-\alpha^2 \bar q^2/2|}.
\end{equation}
Although the integration and the limit $\alpha\rightarrow 0$ in $\alpha F(\omega,q;\alpha)$ do not commute, for zero momentum transfer we find $\alpha F(\omega,\bar q=0)\approx 4 \omega \alpha \log\alpha $ and  in the static limit, $\alpha F(\omega=0)\approx 4 \alpha \bar q$. We present the numerical result $\alpha F(\omega,\bar q)$ and compare the different approximations in Fig.~\ref{fig:anPi}.

\end{document}